\documentclass[10pt,letterpaper]{article}
\usepackage[top=0.85in,left=0.75in,footskip=0.75in,marginparwidth=2.75in]{geometry}

\usepackage[utf8]{inputenc}

\usepackage{cite}
\usepackage[version=3]{mhchem} 
\usepackage{siunitx}
\usepackage{soul}
\usepackage{color}
\usepackage{graphicx}
\usepackage{dcolumn}
\usepackage{bm}

\usepackage{nameref,hyperref}

\usepackage{microtype}
\DisableLigatures[f]{encoding = *, family = * }

\raggedright
\usepackage{changepage}

\usepackage[aboveskip=1pt,labelfont=bf,labelsep=period,singlelinecheck=off]{caption}

\makeatletter
\renewcommand{\@biblabel}[1]{\quad#1.}
\makeatother

\usepackage{lastpage,fancyhdr,graphicx}
\usepackage{epstopdf}
\fancyheadoffset[L]{2.25in}
\fancyfootoffset[L]{2.25in}

\usepackage{color}

\definecolor{Gray}{gray}{.25}

\usepackage{graphicx}

\usepackage{sidecap}

\usepackage{wrapfig}
\usepackage[pscoord]{eso-pic}
\usepackage[fulladjust]{marginnote}
\reversemarginpar

\begin{document}
\vspace*{0.35in}

\begin{flushleft}
{\Large
\textbf\newline{Quantum Transport in Two-Dimensional WS$_2$ with High-Efficiency Carrier Injection through Indium Alloy Contacts}
}
\newline
\\
Chit Siong Lau\textsuperscript{1,$\dagger$},
Jing Yee Chee\textsuperscript{1,$\dagger$},
Yee Sin Ang\textsuperscript{2},
Shi Wun Tong\textsuperscript{1},
LiemaoCao\textsuperscript{2,3},
Zi-En Ooi\textsuperscript{1},
Tong Wang\textsuperscript{1},
Lay Kee Ang\textsuperscript{2},
Yan Wang\textsuperscript{4},
Manish Chhowalla\textsuperscript{4},
Kuan Eng Johnson Goh\textsuperscript{1,5,*}

\bigskip
\bf{1} Institute of Materials Research and Engineering (IMRE), Agency for Science, Technology and Research (A*STAR), 2 Fusionopolis Way, 138634, Singapore
\\
\bf{2} Science, Math and Technology (SMT),
Singapore University of Technology and Design, Singapore 487372
\\
\bf{3} College of Physics and Electronic Engineering, Hengyang Normal University, Hengyang 421002, China
\\
\bf{4} Materials Science \& Metallurgy, University of Cambridge, 27 Charles Babbage Road, Cambridge CB3 0FS, UK
\\
\bf{5} Department of Physics, National University of Singapore, 2 Science Drive 3, 117551, Singapore
\\
\bf{$\dagger$} These authors contributed equally to this work.
\\
\bigskip
* kejgoh@yahoo.com

\end{flushleft}

\section*{Abstract}
Two-dimensional transition metal dichalcogenides (TMDCs) have properties attractive for optoelectronic and quantum applications. A crucial element for devices is the metal-semiconductor interface. However, high contact resistances have hindered progress. Quantum transport studies are scant as low-quality contacts are intractable at cryogenic temperatures. Here, temperature-dependent transfer length measurements are performed on chemical vapour deposition grown single-layer and bilayer WS$_2$ devices with indium alloy contacts. The devices exhibit low contact resistances and Schottky barrier heights ($\sim$10 k$\Omega$\si{\micro\metre} at 3 K and 1.7 meV). Efficient carrier injection enables high carrier mobilities ($\sim$190 cm$^2$V$^{-1}$s$^{-1}$) and observation of resonant tunnelling. Density functional theory calculations provide insights into quantum transport and properties of the WS$_2$-indium interface. Our results reveal significant advances towards high-performance WS$_2$ devices using indium alloy contacts.

{\textbf{Keywords:} 2D materials, contacts, quantum transport, WS$_2$, transition metal dichalcogenides}


\section*{Introduction}
Two-dimensional transition metal dichalcogenides (TMDCs) have been widely studied for their exceptional mechanical and optoelectronic properties, with most of the attention centered on single-layer MoS$_2$.\cite{Chhowalla2016a,Tan2017b,Mak2010b,Radisavljevic2011a,Splendiani2010b} Another intriguing member of the TMDC family is WS$_2$.\cite{Ovchinnikov2014b,Aji2017,Jo2014b,Iqbal2015,Cui2015b,Liu2014d,Sheng2019,Tan2017c,SikHwang2012} Theoretical calculations suggest that WS$_2$ is expected to have higher carrier mobility due to its lower effective mass compared to MoS$_2$, raising interesting prospects for device applications.\cite{Zhang2014h,Kormanyos2014} Yet, significantly fewer studies have been made relative to MoS$_2$. 

It can be challenging to achieve effective contacts in CVD WS$_2$ as the conduction band edge in WS$_2$ is located at a higher energy compared to MoS$_2$, which can result in larger Schottky barriers.\cite{Kim2015b,Aji2017} In particular, quantum transport studies in WS$_2$ have been limited due to the lack of high quality devices with good contacts at cryogenic temperatures.\cite{Jena2014,Allain2015b,Schulman2018}  While promising strategies for room temperature contacts to 2D TMDCs have recently been realized,\cite{Liu2018,Xu2016a,English2016,Das2013a,Guimaraes2016,Liu2015g,Wang2019f} the viability of such contacts at cryogenic temperatures down to 4 K and the understanding of charge injection and transport remain elusive. This poses a major obstacle for the development of quantum devices in 2D TMDCs.

Furthermore, most reports on WS$_2$ devices show large variance in number of layers, preparation methods, \emph{i.e.} exfoliated \emph{vs} CVD, and with poorly defined device geometries, leading to inconsistent results. Understanding low-temperature carrier injection and transport is thus of critical importance for better design and control of TMDC based quantum devices. This will require characterization of scalable devices with properly defined geometries and high quality contacts operable at cryogenic temperatures. 

Here, we present our work on Transfer Length Method (TLM) devices fabricated on single-layer and bilayer CVD grown WS$_2$ using indium alloy contacts. We extract key parameters including contact resistances and Schottky barrier heights, and identify the carrier transport and injection regimes for different temperature and bias ranges. Importantly, our experimental data and Density Functional Theory (DFT) calculations allow us to uncover valuable insights into the nature of low temperature transport and carrier injection in CVD grown single-layer and bilayer WS$_2$.

\section*{Results and discussion}
\subsection*{Field-Effect Transport and Carrier Mobility}

Figure 1a shows an optical image of our fabricated CVD single-layer (SL) and bilayer (BL) WS$_2$ TLM devices. The SL WS$_2$ device schematic and crystal structure of BL WS$_2$ are shown in Figure 1b. Atomic force micrograph is shown in the Supporting Information (Figure S1). 

\begin{figure*}
\centering
\includegraphics[width=15cm]{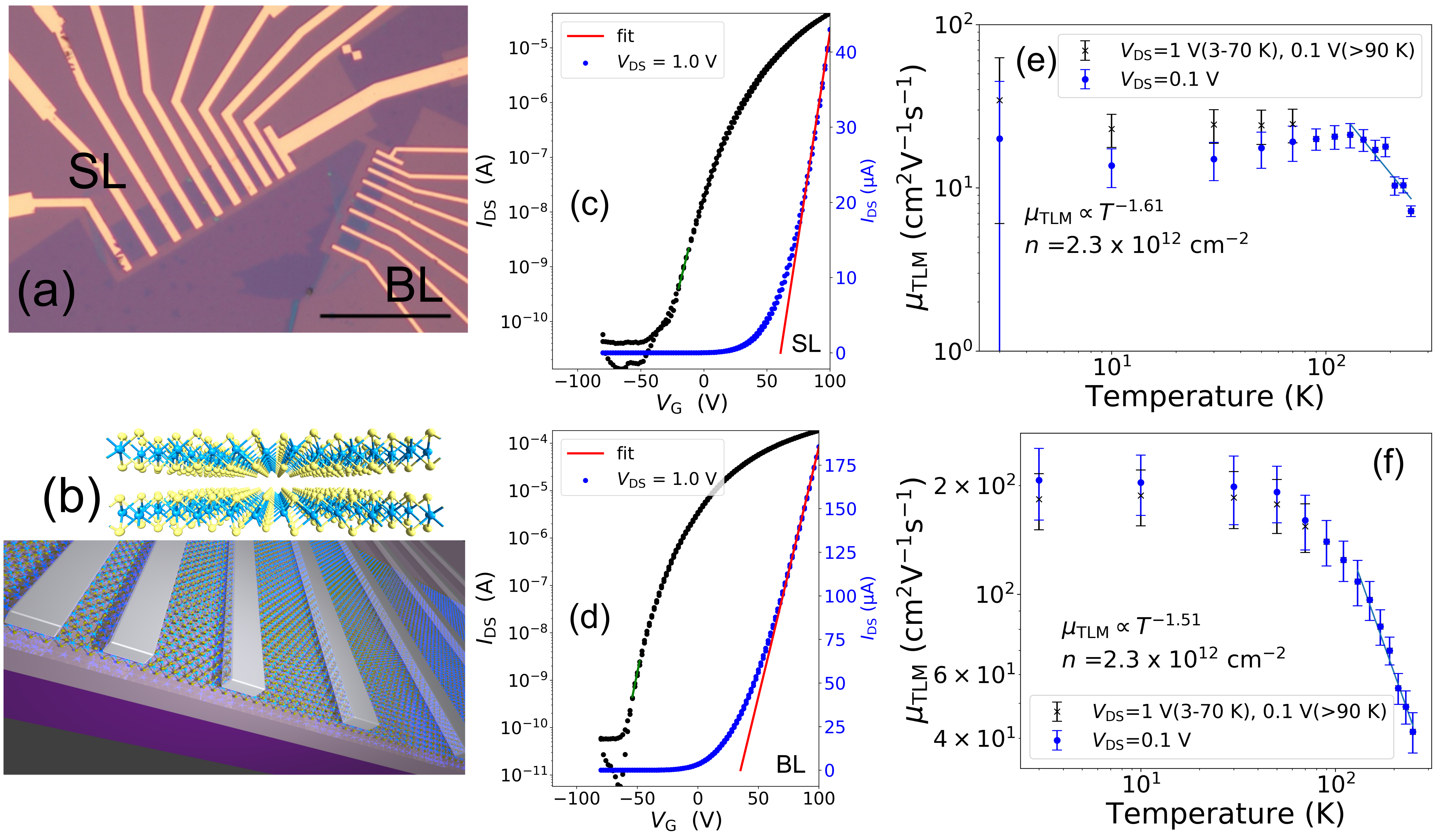}
\caption{a) Optical image of the WS$_2$ single-layer (SL) and bilayer (BL) devices with channel widths 10 \si{\micro\metre} and 5 \si{\micro\metre} respectively. (b) Schematics of single-layer WS$_2$ device and crystal structure of bilayer WS$_2$. Transfer curves for the WS$_2$ (c) single-layer ($L$ = 2 \si{\micro\metre}, $W$ = 10 \si{\micro\metre}) and (d) bilayer devices ($L$ = 1 \si{\micro\metre}, $W$ = 5 \si{\micro\metre}) on linear (blue, right) and logarithmic (black, left) scales. The solid red lines represent the linear fits to the transconductances for extracting $V_\mathrm{TH}$ and $\mu_\mathrm{FE}$. Transfer length method measurements of device resistance as a function of length for the (e) single-layer and (f) bilayer devices. The solid blue lines represent the linear fits to the data.  
}  
\end{figure*}

We determine the electrical quality of the indium alloy contacts and our WS$_2$ devices through field effect and TLM measurements. Our devices exhibit typical \emph{n-}type semiconducting behavior with high on/off ratios ($>$10$^6$/10$^7$ for SL/BL), as estimated from the transfer curves measured at 10 K (Figure 1c and 1d). We also show in the Supporting Information (Figure S6) measurements of our control device where the WS$_2$ was grown from the same batch as the manuscript device and processed in parallel, but with Ti/Au contacts instead of In/Au.

While a widely used technique in literature, the field effect mobility $\mu_{\mathrm{FE}}$ is an underestimation as it ignores the contribution from the contact resistance $R_{\mathrm{C}}$. A more accurate value for the carrier mobility can be extracted using TLM measurements.\cite{Zhong2015} TLM is also the standard method for determining the contact resistance $R_\mathrm{C}$, related to the channel resistance $R_\mathrm{CH}$ and sheet resistance $R_\mathrm{S}$ by the following equation:
\begin{equation}
    R_\mathrm{CH}(L) = R_\mathrm{S}\frac{L}{W} +\frac{2R_\mathrm{C}}{W}. 
\end{equation}
A linear fit of device $R_\mathrm{CH}$ with lengths $L$ at specific net gate voltages $V_{\mathrm{G}}-V_{\mathrm{TH}}$, \emph{i.e}. specific carrier densities $n$, yields the width $W$ normalized $R_\mathrm{S}$ and 2$R_\mathrm{C}$ from the slope and y-intercept respectively (Supporting Information Figure S2). From the sheet resistance $R_\mathrm{S}$, we calculate the carrier mobility $\mu_\mathrm{TLM}$ using:
\begin{equation}
    \mu_\mathrm{TLM} = \frac{1}{qn(V_{\mathrm{G}}-V_{\mathrm{TH}})R_\mathrm{S}(V_{\mathrm{G}}-V_{\mathrm{TH}})}.
\end{equation}
We show the temperature dependence of $\mu_\mathrm{TLM}$ for SL and BL in Figure 2c,d, plotted for source-drain voltages $V_\mathrm{DS}$ = 0.1 and 1 V. The measured $\mu_\mathrm{TLM}$ at 10 K ($V_\mathrm{DS}$ = 1 V) for SL (23 $\pm$ 5 cm$^2$V$^{-1}$s$^{-1}$) and BL (187 $\pm$ 33 cm$^2$V$^{-1}$s$^{-1}$) are among some of the highest reported values for CVD grown WS$_2$. These values are comparable even to devices made from exfoliated flakes, highlighting the excellent quality of our CVD grown material.  
At higher temperatures $>$200 K, $\mu_\mathrm{TLM}$ scales as $T^{-\gamma}$, with $\gamma$ = 1.61 (SL) and $\gamma$ = 1.51 (BL), consistent with reports for other WS$_2$ and MoS$_2$ devices. The values are close to those predicted for MoS$_2$, where $\gamma\approx$ 1.52 (SL) and 2.6 (bulk) and is commonly ascribed to the quenching of the homopolar phonon mode.\cite{Radisavljevic2013,Liu2014d} At lower temperatures $<$70 K, the behaviour of $\mu_\mathrm{TLM}$ differs between the SL and BL devices. For SL, $\mu_\mathrm{TLM}$ decreases with decreasing temperature at $V_\mathrm{DS}$= 0.1 V but for $V_\mathrm{DS}$= 1 V, $\mu_\mathrm{TLM}$ instead saturates at low temperatures. For the BL, $V_\mathrm{DS}$= 0.1 V and 1 V both lead to consistent saturation values for $\mu_\mathrm{TLM}$ at low temperatures. 

\subsection*{Variable Range Hopping and Disorder}

\begin{figure*}
\centering
\includegraphics[width=15cm]{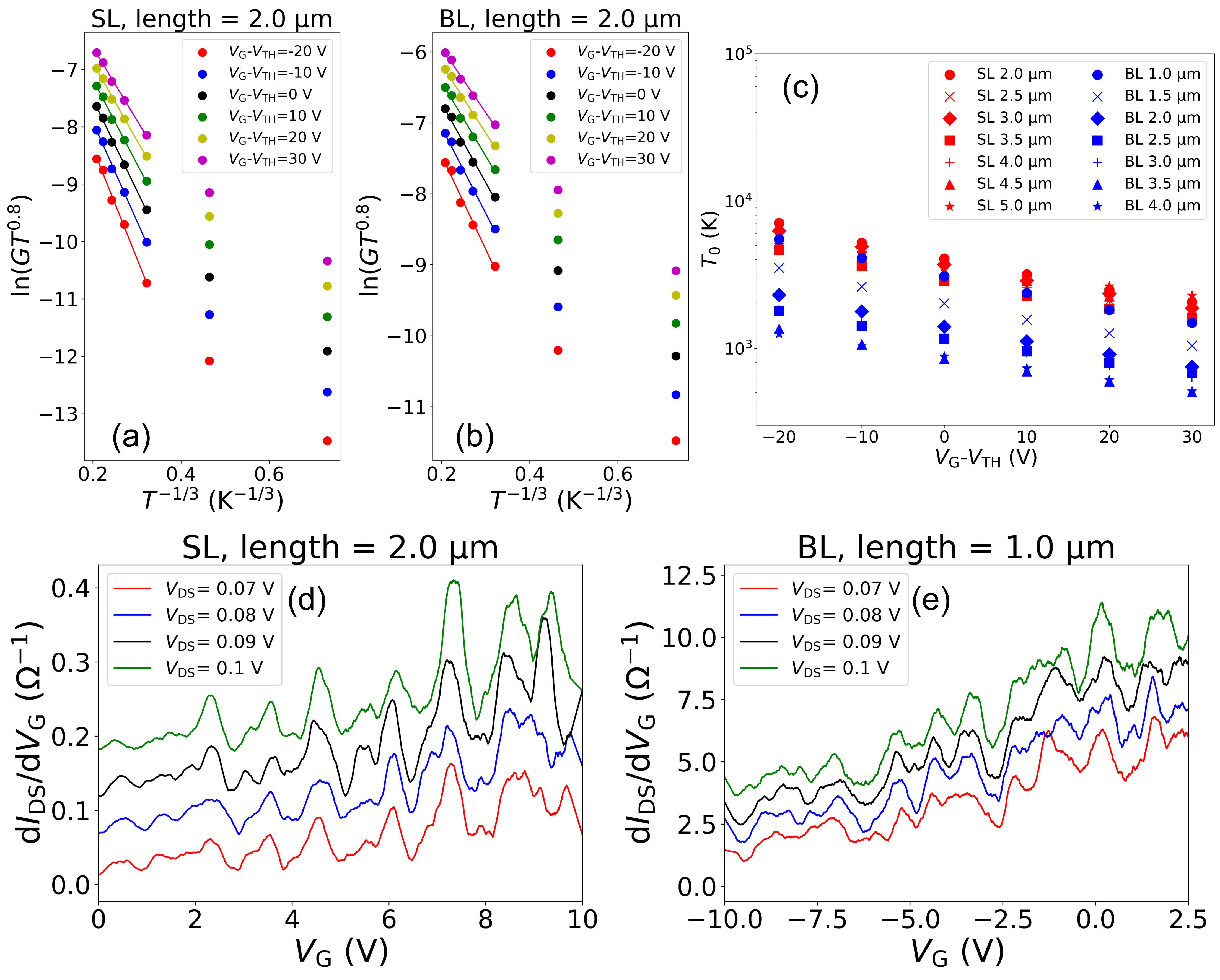}
\caption {Fits to the Variable Range Hopping (VRH) transport model for (a) single-layer and (b) bilayer devices. At lower temperatures below 30 K, the data deviates from the model, indicating the onset of resonant tunnelling through localized states. (c) Extracted correlation energy scale $T_\mathrm{0}$ for single-layer (red) and bilayer (blue) devices. We find little variation between the single-layer devices with different lengths, while $T_\mathrm{0}$ decreases with increasing device lengths for bilayer. High resolution differential conductance curves taken at 3 K for (d) single-layer ($L$=2 \si{\micro\metre}) and (e) bilayer ($L$=1 \si{\micro\metre}). Reproducible peaks suggesting resonant tunnelling through localized states are clearly visible in both devices across a range of $V_\mathrm{DS}$. }
\end{figure*}

While mobility degradation at low temperatures is commonly interpreted as due to scattering from charge impurities,\mbox{\cite{Radisavljevic2013}} it could also be due to the invalidity of applying the mobility equation since the device is operating at a non-linear region. Another possible origin is the underestimation of mobility due to Schottky barriers at small V$_\mathrm{DS}$.\mbox{\cite{Liu2014d}} At small $V_\mathrm{DS}$, the mobility can be underestimated due to Schottky barriers, whereas contact resistance becomes negligible when $V_\mathrm{DS}$ is increased. The saturation of $\mu_\mathrm{TLM}$ ($V_\mathrm{DS}$= 1 V) at low temperatures for both the SL and BL devices point to effective damping of the Coulomb scattering from charged impurities.\cite{Radisavljevic2013} To better understand the nature of the low-temperature transport, we use the variable range hopping (VRH) model:
\begin{equation}
    G = G_\mathrm{0}(T)\exp(-T_\mathrm{0}/T)^{1/3}.
\end{equation}
Here, $G_\mathrm{0}$ is the conductance, $T_\mathrm{0}$ the correlation energy scale and $G_\mathrm{0}$ = $AT^m$ with $m\approx$ 0.8-1.\cite{Ghatak2011b} Figure 2 shows the fits to our data at different $V_\mathrm{G}-V_\mathrm{TH}$ 
for the (a) SL and (b) BL devices. We find two distinct regimes, (1) 30 $<T<$ 130 K and (2) $T<$ 30 K. 

For regime (1), the data agrees very well with the VRH transport model, suggesting that transport in WS$_2$ occurs over a wide band of localized states.\cite{Ghatak2011b,Radisavljevic2013} For regime (2), the weakening of $G$ with $T$ points to the transition towards a strongly localized regime. Such transport behaviour in 2D materials was previously attributed to the presence of a disordered potential landscape.\cite{Ghatak2011b,Haneef2020} Indeed, the high resolution differential conductance curves taken at 3 K, shown in Figure 2d (SL) and 2e (BL), exhibit stable reproducible peaks across a range of $V_\mathrm{DS}$. These peaks are indicative of resonant quantum tunnelling through localized states in a disordered potential landscape, consistent with the low temperature behaviour of our VRH model in Figure 2a,b.

The extracted $T_\mathrm{0}$ values are shown in Figure 2c. We find that $T_\mathrm{0}$ decreases with increasing carrier densities, expected for strongly localized 2D electron systems where the Fermi energy lies in the conduction band tail.\cite{Pepper1974,Ghosh2002} Notably, values of  $T_\mathrm{0}$ for all 14 devices lie within a single order of magnitude. This strongly suggests that the origin of the disordered potential landscape is extrinsic, for instance carrier traps in the substrate/interface, rather than intrinsic, such as structural defects/chemical impurities.\cite{Haneef2020,Ghatak2011b} A disordered potential landscape is consistent with the larger SL $T_\mathrm{0}$ values as SL devices are more strongly influenced by substrate induced disorder compared to BL devices. The higher variance for $T_\mathrm{0}$ in BL devices suggests that intrinsic disorder such as random impurities and defects plays a more prominent role compared to substrate disorder. This implies that transport in the BL devices occur primarily through the indium-contacted top layer with partial screening from the bottom layer, consistent with our first-principle DFT calculations (discussed below).

Furthermore, we find that $T_\mathrm{0}$ generally decreases with increasing length for BL devices, suggesting a greater density of defects in the shorter channel devices. Considering that the contact regions constitute a greater proportion of the system for shorter channel devices, this indicates that the bulk of the disorder lies close to the metal-semiconductor interfaces. Indeed, such disorder at the metal/semiconductor interface was shown to be caused by lattice damage during metal deposition or increased polymer residue from resist development during fabrication.\cite{Wang2019f,Liu2018} The further reduction of evaporation rate during metal deposition can mitigate these destructive effects.\cite{Wang2019f}

\subsection*{Temperature Dependence of Contact Resistance and Schottky Barrier}

\begin{figure*}
\centering
\includegraphics[width=15cm]{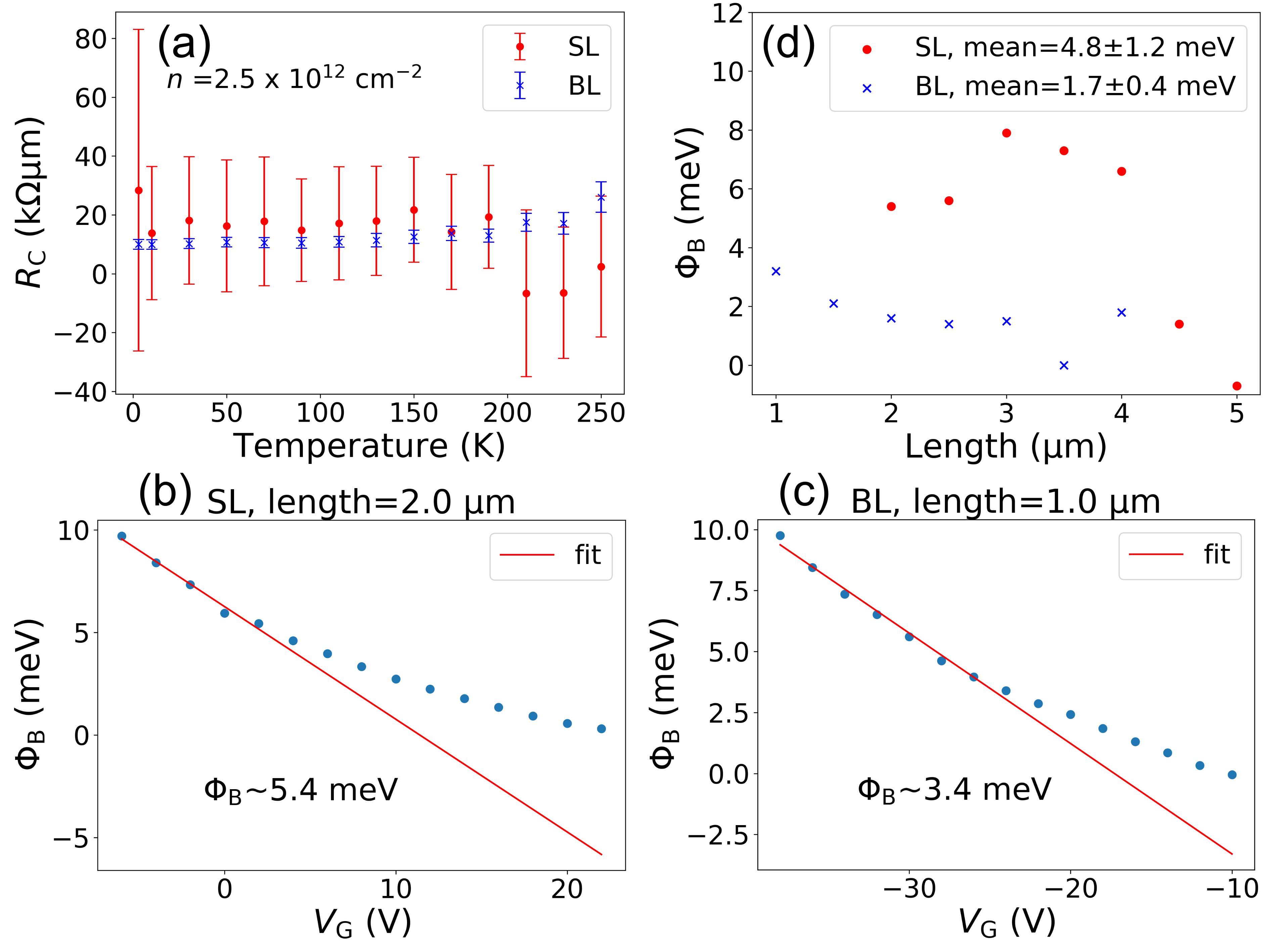}
\caption {(a) Temperature dependence of the contact resistances $R_\mathrm{C}$. Schottky barrier heights $\Phi_\mathrm{B}$ \emph{vs} gate voltage $V_\mathrm{G}$ for (b) single-layer (length= 2.0 \si{\micro\metre}) and (c) bilayer (length= 1.0 \si{\micro\metre}) as determined from the Richardson plots (Supporting Information Figure S3). The gate voltage where $\Phi_\mathrm{B}$ tends away from the linear fit is the flatband condition for which the true Schottky barrier height is determined. (d) Schottky barrier heights $\Phi_\mathrm{B}$ for the different device lengths.
} 
\end{figure*}

Having established the nature of low temperature transport through our devices, we now discuss the contacts. We estimate the influence of the width normalized contact resistances $R_\mathrm{C}$ through the TLM measurements using Equation 2. The temperature dependence of $R_\mathrm{C}$ is shown in Figure 3a. We find that $R_\mathrm{C}$ tends to saturate at low temperatures for both SL and BL devices, with $R_\mathrm{C}$ = 28 $\pm$ 45 (SL) and 9.9 $\pm$ 1.6 (BL) k$\Omega$\si{\micro\metre} at 10 K (Supporting Information Figure S2). These values compare favorably with literature reported values, indicating the excellent metal-semiconductor contact between indium alloy and WS$_2$.\cite{Allain2015b,Liu2015g,Wang2019f,Das2013a,Park2016b,Guimaraes2016} For a metal/semiconductor contact with sizable Schottky barrier height, the charge injection is dominated by thermionic emission and the contact resistance decreases exponentially with temperature as $R_\mathrm{C} \propto \exp\left(\Phi_\mathrm{B}/k_\mathrm{B}T\right)$. The weak temperature dependence of $R_\mathrm{C}$ in our devices suggests that charge injection across the metal/semiconductor contact is not limited by thermionic emission, and the Schottky barrier height is expected to be small.

To further assess the nature of our contacts, we extract the Schottky barrier heights $\Phi_\mathrm{B}$ of our devices by measuring the activation energy in the thermionic emission region at low gate voltages. Here the current $I_\mathrm{DS}$ is determined by:
\begin{equation}
    I_\mathrm{DS} = A^*T^{3/2}\exp\left(\frac{-\Phi_\mathrm{B}}{k_\mathrm{B}T}\right)(1-\exp\left(\frac{-qV_\mathrm{DS}}{k_\mathrm{B}T}\right),
\end{equation}
where $A^*$ is the Richardson constant, $V_\mathrm{DS}$ the bias, $T$ the temperature and $k_\mathrm{B}$ the Boltzmann's constant. With this equation, the Richarson plot, ln($I_\mathrm{DS}/T^{3/2}$) $\approx$ 1/$T$ (Supporting Information Figure S3), yields $\Phi_\mathrm{B}$ as a function of gate voltage which is shown in Figure 3b (SL) and 3c (BL). The true $\Phi_\mathrm{B}$ is extracted using the flat band voltage condition, determined as the gate voltage at which $\Phi_\mathrm{B}$ tends away from the linear fit, and shown in Figure 3d for all device lengths. We find that the BL devices generally exhibit lower Schottky barrier heights, with $\Phi_\mathrm{B}$ = (BL) 1.7 $\pm$ 0.4 and (SL) 4.8 $\pm$ 1.2 meV. Notably, the extracted Schottky barrier heights are at least an order of magnitude less than what are typically reported in literature using other contact metals ($>$100 meV).\cite{Zheng2019,Kaushik2014,Fan2016a,Shih2014} The low $\Phi_\mathrm{B}$ values further confirm the quality of our indium alloy contacts and are consistent with our interpretation of the mobility $\mu_\mathrm{TLM}$ curves (Figure 2c,d). At $\Phi_\mathrm{B}\approx$ 6 meV, the corresponding thermal energy translates to a temperature of $\approx$ 70 K, close to the temperature at which $\mu_\mathrm{TLM}$ begins to diverge for different $V_\mathrm{DS}$ in the case of the SL (Supporting Information Figure S4). For the BL device, the lower $\Phi_\mathrm{B}$ and $R_\mathrm{C}$ at low temperatures compared to the SL device result in little difference between the $\mu_\mathrm{TLM}$ curves for $V_\mathrm{DS}$ = 0.1 V and 1 V. 

We now discuss the relation between Schottky barrier heights and $R_\mathrm{C}$ in our devices. The contact resistance of a metal/2D-semconductor interface is not solely determined by the metal/semiconductor Schottky barrier, but a combination of (i) thermionic charge injection over Schottky barrier heights; and (ii) carrier scattering in the 2D semiconductor under the metal contact region.\mbox{\cite{English2016}} As the Schottky barrier heights exhibit an exceptionally low value, the contact resistance is expected to be dominated by the carrier scattering in the metal contacted WS$_2$ region, rather than by thermionic carrier injection across the metal/semiconductor interface. The temperature dependence of $R_\mathrm{C}$ offers an obvious signature to distinguish the charge injection and transport mechanism across a Schottky contact. In thermionic-limited transport, $R_\mathrm{C}$ is expected to decrease rapidly with temperature as thermionic charge injection is approximately exponentially promoted at elevated temperatures. However, such strong temperature dependence of $R_\mathrm{C}$ is absent in our device (Figure 3a),  suggesting that thermionic charge injection is not a limiting transport process across the metal/semiconductor contacts, in agreement with the measured ultra-low Schottky barrier heights (Figures 3c and 3d). As demonstrated in our DFT simulations (Figure 5), the strong metallization between In and WS$_2$ can lead to the generation of substantial defect sites that leads to higher $R_\mathrm{C}$. We suspect that these carrier scattering effects are the dominant contributor of $R_\mathrm{C}$ in our devices.

\subsection*{Charge Injection Mechanism at Indium/WS$_2$ Interface}

We now investigate the charge injection and transport mechanisms in the SL and BL devices. Across a metal/semiconductor contact, charge injection occurs \emph{via} three distinct pathways (see Figure 4a-c): (i) direct tunneling (DT) at low $V_\mathrm{DS}$;\cite{Simmons1963} (ii) Richardson-Schottky (RS) thermionic injection at moderate $V_\mathrm{DS}$;\cite{Schottky1938, Crowell1969} and (iii) field-induced Fowler-Nordheim (FN) tunneling at high $V_\mathrm{DS}$ \cite{Murphy1956} where the current-voltage scaling relations are governed, respectively, by 

\begin{equation}
	\ln\left(\frac{I_\mathrm{DS}}{V_\mathrm{DS}^2}\right)\propto\ln\left(\frac{1}{V_\mathrm{DS}} \right) \& \text{ , DT}
\end{equation}

\begin{equation}
	\ln\left(I_\mathrm{DS}\right) \propto V_\mathrm{DS}^{1/2} \& \text{ , RS emission}
\end{equation}

\begin{equation}
	\ln\left(\frac{I_\mathrm{DS}}{V_\mathrm{DS}^2}\right) \propto -\frac{1}{V_\mathrm{DS}} \& \text{ , FN tunneling}
\end{equation}

\begin{figure*}
\centering
\includegraphics[width=15cm]{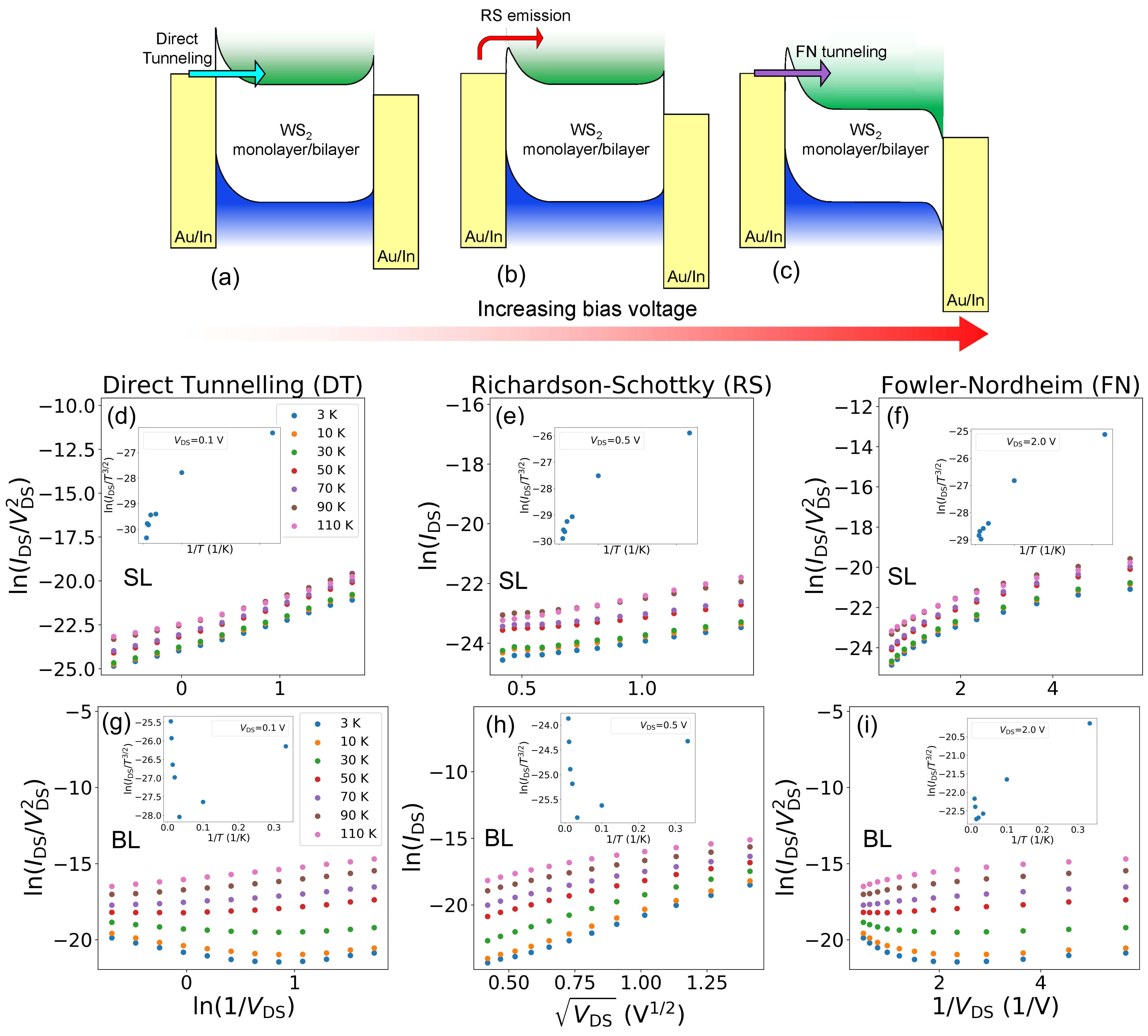}
\caption {Schematics showing the charge injection mechanisms for (a) direct tunnelling (DT), (b) Richardson-Schottky (RS) emission and (c) Fowler-Nordheim (FN) tunnelling. Current signatures for the different models for (d-f) single-layer (g-i) and bilayer  devices. Insets show the thermionic Arrhenius plots In($I_\mathrm{DS}/T^{3/2}) \propto$ -1/$T$ at the different bias regimes. Full-sized figures of the insets are available in the Supporting Information (Figure S5).}
\end{figure*}
In Figure 4d-g, we show the DT, RS and FN plots for the SL (length = 2 \si{\micro\metre}) and BL (length = 2 \si{\micro\metre}) devices at different temperatures and $V_\mathrm{G}$ = -50 V (Supporting Information Figure S7 shows ln($I_\mathrm{DS}$) \emph{vs} ln($V_\mathrm{DS}$) plots). These SL and BL devices exhibit contrasting transport behavior. For the SL device, the best linear fit for the ohmic current-voltage relation in Equation 5 (Figure 4d) spans the entire measured bias voltage range. While there is some linear behavior for the RS plot at intermediate to high bias voltages (Figure 4e), the thermionic Arrhenius plots of In($I_\mathrm{DS}/T^{3/2}) \propto$ -1/$T$ (insets of Figure 5d-f) do not exhibit any negative slope trends. This rules out interfacial thermionic-based injection as a dominant transport mechanism over the temperature range 3 K $\leq T\leq$ 110 K.

Charge transport across the device is determined by both the charge injection across the contact and the charge conduction in the bulk. Because of the ultra-low Schottky barrier heights in SL device, the lower $\mu_\mathrm{TLM}$ compared to BL suggests that transport in SL device is more strongly influenced by the channel, compared to the contact. As charge conduction across the SL channel also exhibits ohmic behavior (\emph{i.e}. the bulk current $I_\mathrm{bulk} \approx ne\mu V_s / L$) with the same current-voltage scaling relation as Equation 5, the conduction in the SL device is likely \emph{bulk-limited} rather than \emph{injection-limited}. This is further supported by the absence of FN tunneling in the high-bias regime of the FN plot in Figure 4f.

In contrast, the BL device (Figure 4g-i) exhibits \emph{injection-limited} conduction. Due to the substantially higher $\mu_\mathrm{TLM} \approx 187$ cm$^2$V$^{-1}$s$^{-1}$ in BL device (compared to $\mu_\mathrm{TLM} \approx 23$ cm$^2$V$^{-1}$s$^{-1}$ for SL device), charge injection across the contact plays a relatively more prominent role in limiting the device output characteristics. The injection-limited nature of the charge transport in BL device is confirmed by the thermionic Arrhenius plot which exhibits a negative slope at temperature $T\ge$ 30 K, a signature of thermionic charge injection across the metal/semiconductor interface (Figure 4g).

From small to intermediate bias voltages, linear behavior is observed for both the DT and RS models (right region of Figure 4g and left region of Figure 4h). At higher bias voltages, the DT plots deviates from linear behaviour as the charge injection mechanism transits towards FN emission, confirmed by the signature negative slopes of the FN model (Figure 4i). With increasing temperatures at $T>$ 50 K, we find that the FN plot no longer exhibits a negative slope, indicating the dominance of thermionic emission. This is further corroborated by the thermionic Arrhenius plot at high-bias regime, which exhibits the signature negative-slope trend at $T>$ 50 K (inset of Figure 4i). 

The charge injection and transport mechanisms in Figure 4 thus reveal fundamentally different charge transport mechanisms between the SL and BL devices, despite both having similarly low values of $R_\mathrm{C}$ and $\Phi_\mathrm{B}$.

\subsection*{Density Functional Theory Simulation}

\begin{figure*}
\centering
	\includegraphics[width=15cm]{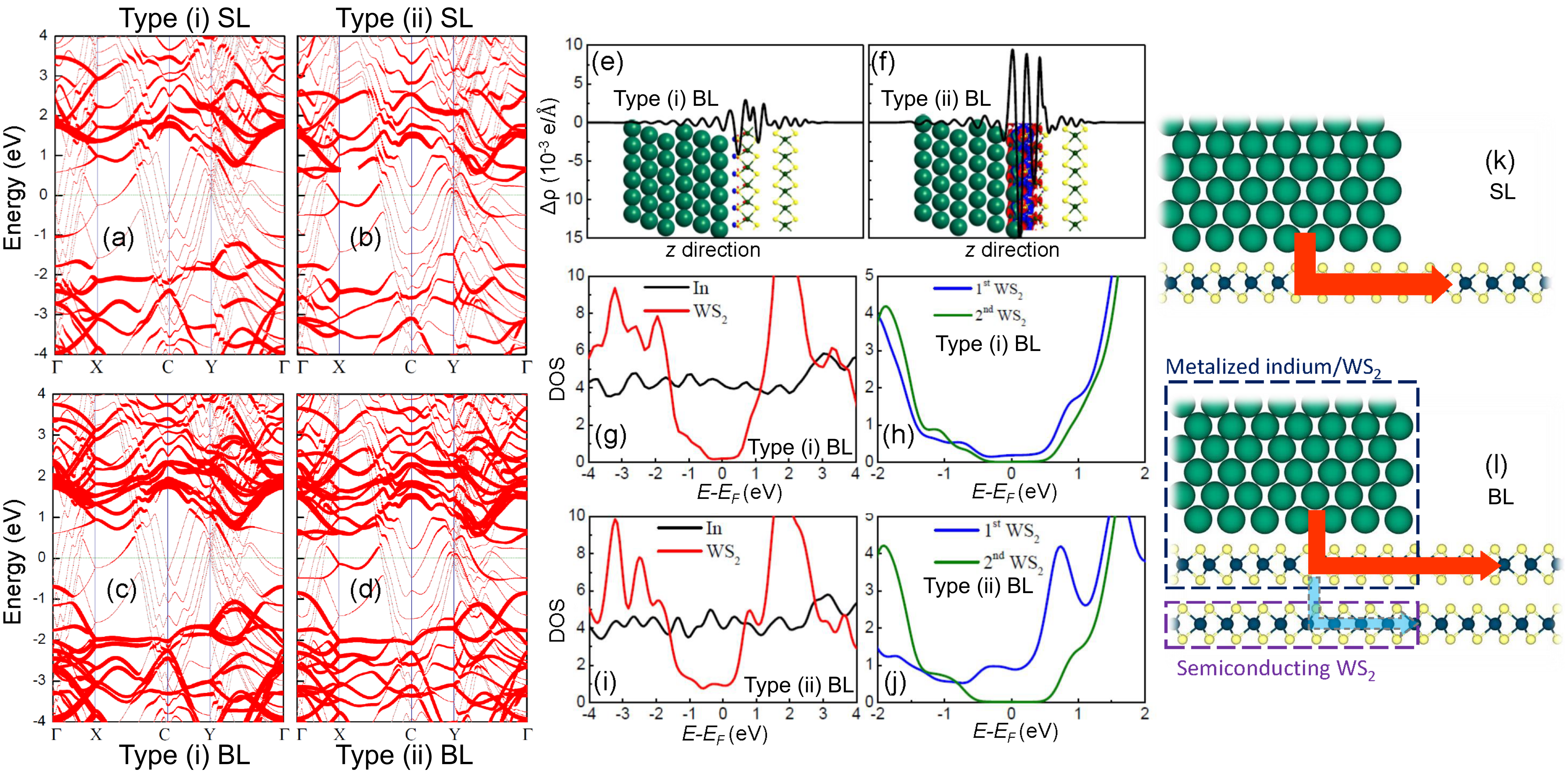}
	\caption{Density functional simulations of indium/WS$_2$ contact. Projected band structure of (a,b)/(c,d) single-layer(SL)/bilayer(BL) WS$_2$ contacted by indium with (a,c) type (i) contact of indium/WS$_2$ (separation $d$ = 2.99 \si{\angstrom} (SL) and $d$ = 2.98 \si{\angstrom} (BL)); and (b,d) type (ii) contact of indium/WS$_2$ (separation $d$ = $1.68$ \si{\angstrom}). Charge transfer, $\Delta\rho$ in bilayer WS$_2$ with (e) type (i) and (f) type (ii) contact. Calculated partial density of states (DOS) of the (g)/(i) type (i)/type(ii) bilayer WS$_2$/indium contact, and the (h)/(j) type (i)/type(ii) top layer indium-contacted WS$_2$ (denoted as 1st WS$_2$) and the bottom layer WS$_2$ (denoted as 2nd WS$_2$). Schematic drawings of the charge injection mechanisms across the (k) single-layer and (i) bilayer WS$_2$/indium contacts.}
\end{figure*}

To validate our experimental findings and gain further insights on the electronic properties of the WS$_2$-indium contact, we turned to DFT calculations (Figure 5). Because the WS$_2$ devices are fabricated using conventional deposition of indium, the metal/2D-semiconductor interface can be non-ideal and more prone to defects, strain and chemical bonding.\cite{Liu2018} For completeness, we performed DFT calculations on indium-contacted WS$_2$ SL and BL with two distinct contact types \cite{kong}: (i) `clean' contact with an optimized indium/WS$_2$ corresponding to an atomically-sharp Van der Waals metal/2D-semiconductor interface fabricated \emph{via} mechanical transfer of metal onto 2D semiconductor \cite{Liu2018}; and (ii) covalent-bonded contact corresponding to a non-ideal contact fabricated using conventional methods of metal deposition. DFT calculations for MoS$_2$/indium are also available in the Supporting Information (Figure S8). 

From the band structure diagrams in Figure 5a-d, we find substantial \emph{metallization} in both type (i) and (ii) contacts. The overlap of WS$_2$ and indium electronic states around the Fermi level reveals the absence of Schottky barriers at the indium/WS$_2$ interfaces. This is consistent with our exceptionally low experimental values of $\Phi_\mathrm{B}$ ($<10$ meV). The nonzero $\Phi_\mathrm{B}$ measured in our devices is likely to originate from imperfections and defects at the indium/WS$_2$ interface. 

In Figure 5e and 5f, the calculated charge density variation, $\Delta \rho$, of the bilayer-WS$_2$/indium interface indicates significant amount of charges transferred from indium into metal-contacted top layer WS$_2$. We thus expect the metal-contacted top layer WS$_2$ to play an importantly role in both the charge injection and transport in our devices.

To elucidate the layer-contrasting metallization and transport, we calculate separately the density of states (DOS) of indium and WS$_2$ (Figure 5g), and the top layer (indium-contacted) WS$_2$ and the bottom layer WS$_2$ (Figure 5h), for type (i) and (ii) (see Figure 5i and 5j) contacts. By inspecting the DOS around the Fermi level, the bottom layer WS$_2$ remains semiconducting, while the metal-contacted top layer becomes highly metallic. 

Importantly, the top and the bottom WS$_2$ is separated by a large transport barrier of about 0.4 eV (Figure 5j and 5h). We thus expect that the domination of the top-layer carriers is not limited to carrier injection at the metal/semiconductor interface, but that subsequent carrier transport in the 2D channel is also concentrated in the top indium-contacted WS$_2$ layer. This effect is further amplified at low temperatures where inter-layer thermal-assisted carrier injection and phonon-assisted scattering are strongly suppressed.

The layer-dependent metallization and carrier transport predicted by our DFT simulations are thus in unison with the $T_0$ extracted at the low-temperature regime of $30<T<130$ K (Figure 2c),  where the lower variance of the $T_0$ values for BL device can be attributed to the dominant role of top layer indium-contacted WS$_2$ as the primary carrier transport channel (see Figure 5k and 5l for the schematic drawings of the carrier injection and transport in SL and BL WS$_2$/indium contact, respectively). 

\section*{Conclusion}
To conclude, we performed temperature dependent electrical measurements on single-layer and bilayer CVD grown WS$_2$ transistors with indium alloy contacts. We find that the high quality of the indium alloy contacts persists down to 3 K leading to excellent performance in our devices. We used DFT calculations to unravel hitherto unexplained carrier injection and transport mechanisms. These results represent an important advance for the study of quantum transport in 2D TMDCs and the development of scalable TMDC based quantum devices, where progress had been limited by poor contacts and low quality of scalable TMDC materials.

\section*{Experimental}
\textbf{Device Growth and Fabrication.} We first cleaned the growth substrates (SiO$_2$/Si) with acetone, isopropanol and DI water under 10 min ultrasonication. The pre-cleaned SiO$_2$/Si substrates were then treated with oxygen plasma for 20 seconds in Harrick Plasma Expanded Plasma Cleaner at fixed RF power (30W) and oxygen flow rate of 8 sccm. We next prepared the W-based precursor solution by dissolving 3 mg sodium tungstate dihydrate (Na$_2$WO$_4\cdot$2H$_2$O) in 10 ml DI water. After spin coating the precursor solution on the plasma-treated substrates at 3500 rpm for 40 s, we annealed at 130 $^{\circ}$C for 10 min. We performed the atmospheric pressure growth in a fused quartz tube at 800 $^{\circ}$C under 60 sccm of N$_2$ gas. A crucible filled with 1500 mg of sulfur powder (99.999 \%) was placed upstream and heated to a temperature of 200 $\pm$ 20 $^{\circ}$C. After 10 min growth, we cooled the furnace to room temperature under 300 sccm N$_2$ gas flow.

The grown flakes were wet transferred onto a 290 nm SiO$_2$/Si substrate that also serves as a back-gate for tuning the overall carrier densities in our devices. PMMA was spun onto the surface of the growth SiO$_2$ substrate which was then etched using dilute HF (5\%) to detach the PMMA/WS$_2$ film. We rinsed the film in a series of 3 DI water baths. Finally, the film was transferred onto the target substrate (also dipped in dilute HF (5\%) and rinsed) and dried overnight before PMMA removal with Microposit Remover 1165.
 
Next, we patterned the source/drain electrodes using standard e-beam lithography and subsequently deposited 3/40 nm of In/Au at 0.5/1.0 \si{\angstrom/s}. The channel geometries were then defined using e-beam lithography and etched with SF$_6$ plasma.\cite{Kotekar-Patil2019n} Finally, we annealed the devices at 200 $^{\circ}$C in 10\% H$_2$/Ar forming gas to remove resist residual and further improve the contacts. Transport measurements were performed in a closed-cycle Janis cryostat using Keithley 2450 sourcemeters. 
 
\textbf{Computational Details.} The density functional theory (DFT) calculations were performed by Vienna \emph{Ab Initio} simulation package (VASP). \cite{Kresse,Kresse1} We adopted the Perdew-Burke-Ernzerhof (PBE) functional of the generalized gradient approximation (GGA) to describe the exchange correlation interaction.\cite{Kresse2} The $(1 \times \sqrt 3)$ unit cells of In in (101) orientation were adjusted to match the $(1 \times \sqrt 3)$ unit cell of WS$_2$. The k-point sampling was set to $9 \times 15 \times 1$. The energy cutoff and self-consistent convergence accuracy were chosen at 500 eV and 1 $\times$ 10$^{-6}$. The DFT-D3 method of Grimme was used to describe the interlayer van der Waals interaction.\cite{Grimme} The dipole correction was also added, and the forces on all atoms were less than 0.01 eV/$\si{\angstrom}$ for the structural relaxation. A 15 $\si{\angstrom}$ vacuum layer was chosen along the z direction to avoid the artificial interactions between adjacent layers.

\section*{Supporting Information}
Supporting information is available at http://pubs.acs.org.

\section*{Acknowledgments}
This research was supported by the Agency for Science, Technology and Research (A*STAR) under its A*STAR QTE Grant No. A1685b0005 and CDA Grant No. A1820g0086. Y.S.A., L.C. and L.K.A. acknowledge the supports of Singapore MOE Tier 2 Grant (2018-T2-1-007) and  USA ONRG grant (N62909-19-1-2047). All the calculations were carried out using the computational resources provided by the National Supercomputing Centre (NSCC) Singapore. The authors thank C.H.K. Goh for assistance with indium evaporation in device fabrication.

\bibliography{Indium}

\bibliographystyle{abbrv}

\end{document}